*Ab Initio* Calculation of the Anomalous Acceleration of Pioneer 10 *In Vacuo*

R. Anania[1] & M. Makoid[2]
[1]*Creighton University emeritus, 2500 California Plaza, Omaha NE 68178*
*email anania@creighton.edu*
[2] *Creighton University, 2500 California Plaza, Omaha, NE 68178*
*email makoid@creighton.edu*[2]

**Abstract**

The anomalous acceleration of Pioneer 10 is presented as a calculation using a simple optical model. The model is based on the bending of background gravity behind the Sun in the same way that light is bent by the Sun. Structures of ponderable matter about the Solar system, neutron stars, and galaxies are described. Viewable red and blue shiftings of light are predicted.

**Section 1 :** Introduction

Pioneer 10 and 11 are on trajectories in opposite directions from each other, and which are also perpendicular to a line towards the central core of the galaxy. It has been reported that they both exhibit a similar and nearly equal anomalous acceleration towards the direction of the Sun, which is approximately constant in magnitude, and that for Pioneer 10 is better known (Anderson, 2002).

If the acceleration is indeed from a central force, and while the direction is through the Sun, it need not be directly from the Sun. The model which is presented here for discussion, is that the accumulated force for the anomalous acceleration acting on Pioneer 10, is primarily passing through the Sun and has as its origin, uncancelled and unbalanced sources which are on the other side of the Sun.

The model is optical in nature, and has as its central feature, that the force of gravity is bent as is light by a massive object such as the Sun (Wheeler, 1973). While a ray of light impinging on an object yields a repulsive force, a ray of gravity yields an attractive force. External light is stopped in its path by a massive object such as the Sun, yet external gravity passes on through and is bent in its path by the mass and gravity of the Sun. Papini (Papini, 1991) established, in the weak field limit, that electromagnetism and gravity can both be placed into the same four vector potential. The four vector potential, of course, contains geometrical optics, so gravity and electromagnetism should be treated similarly, optically. Sommerfeld (Sommerfield, 1952) established, in the static limit of electrostatics and magnetostatics, that they follow the standard rules of refraction and reflection of static forces in geometrical optics. Together then, electrostatics, magnetostatics, and gravitostatics all obey the same rules of refraction of forces in geometrical optics, in the weak field and static limits. Note: In the static limit, there is no far field diffraction since all optics are in the near field. The trajectories of the geodesics of massless particles that travel at the speed of light in the weak field limit do not depend on the energies of the particles and thus are bent equally.

---

[2] Corresponding Author

The optical model then is that the Sun is a spherical lens which bends or refracts external gravity, while it is also a source itself, and the totality acts on Pioneer 10. As a lens, it has a focal length, and spherical aberration which is indicative of the density profile of the Solar interior, which may change with time. The Sun images the sources of these gravitational rays of attractive force and they are of the local stellar neighborhood, and the galactic and cosmic background. The analogue of an intermediate near field focal surface is a test particle on a trajectory, a probe such as Pioneer 10. The departure from Celestial Mechanics by imaging is indicative of the change in the force, the anomalous acceleration.

The focal axis is the line through the center of the Sun and the observation point. Inside the focal point, rays reach the axis through the Sun.

An *ab initio* calculation is the essence of this model which is based on first principles, and it allows the determination of the anomalous acceleration of Pioneer 10 as a function of radial distance from the Sun.

The experimental and observational implications (Wheeler & Ford, 2005) of this model are discussed. They then may be partitioned into local and far.

Other objects such as galaxies and neutron stars should show this affect. On a cosmic scale, Olber's paradox may be revisited; why is the sky not uniformly bright with intense light and gravity?

It is in Section 2 that a formula for the force is presented based on the principles of optics. The one free parameter of the external mass is fixed by the data for the motion of Pioneer 10 at an efficacious point. It is fortuitous that the motion is perpendicular to the direction of the galactic core and that the image of the external mass is approximately the same for both Pioneers, and the anomaly is not complicated further by the core.

In section 3, the source of the data (NASA, 2004) and its characteristics are described. The data is compared to three formulae: they are a straight line, the model for an additional constant acceleration, and the optical model. The goodness of fit to the data are compared.

In section 4 observational and experimental implications are discussed with respect to the location of Kuiper belts and Oort clouds around the Sun and neutron stars. The implications of structures for galactic objects are delineated.

There are two models in contention for the anomalous acceleration, a constant acceleration and the optical model. The formulae for the force laws from each model are integrated to potentials, and they are compared.

The potential is inserted into the metric to obtain a non-Schwarzschild term in the weak field limit. This is used to predict color shifts of light on and near the axis as further experimental implications of the optical model.

Section 5 contains final discussions and conclusions.

Section 6 contains references and citations.

<div align="center">

**Section 2**  -  *Ab Initio* Calculation of an Optical Model
of the Bending of Gravity by Gravity

</div>

 As is well known, the bending of light around the Sun has been verified and is in accordance with General Relativity and the Parameterized Post-Newtonian (PPN) formalism. A spin-one ray of particles of light is bent around the Sun along its geodesic by the form of the metric by the geometrical warping of space with zero curvature, independently of its energy or spin, solely on its properties that it is massless and travels at the speed of light. (Lorentz, 1952)

Like rays of photons which are the progenitors of electromagnetic phenomena, the emissaries of the geometrodynamic warping of space are rays of gravitons, massless spin-2 particles that also travel at the speed of light.

Gravity, too, by the same theory, is therefore expected to be bent in exactly the same fashion as is light. As light originating in or outside the Solar system is bent, so also is gravity.

Caveat: many articles and books have been written about the scattering of particles by fields witch are based on various spins. The currently and experimentally verified theory of gravity does not include a detectable way of distinguishing the scattering or bending amongst those particles which are massless.

Ponderable Mass in Classical Mechanics and General Relativity is based on scalar particles of spin zero and natural parity: however, no stable variety is known to exist. Is anything impossible in world that does not exist?

The optical model of the Solar Marble is that it is a source of illumination, and it is transparent to rays passing through it. Latter rays passing around it will be considered.

The illumination, I, is the sum of three terms. The first is the divergence of the Solar source, the second is the convergence of rays from the background that have passed through the Sun, and the third is the umbra of the Sun. The first is x, the latter two are y, and their ratio, s, is sensitive to the background brightness.

$$I = x + y \qquad\qquad (2.1)$$

$$s = \frac{y}{x} \tag{2.2}$$

$$x = \frac{A}{r^2} \tag{2.3}$$

$$y = B \left[ \frac{1}{(f-r)^2} - \frac{1}{f^2} \right] \tag{2.4}$$

r is the on axis distance from the center of the sun, and f is its focal length, and for now spherical aberration is ignored. A and B are defined below. There is reason to later express B as A∘R subsequently.

Although the illumination is treated as a scalar, its on axis value is a sum of vectors with angles of inclination less than a degree of arc, and so a cosine term for projection onto the optical axis is not required. There are two values of r that are representative of an optical model, y disappears where r is zero and at twice f. At zero, convergence has not begun and there is no umbra. At 2f the rays have passed the focal point and have become divergent, filling in the umbra. Beyond 2f, the umbra persists.

If this process is imagined as progressing from the left to the right, then in the case of an isotropic background, there also is light from the right reaching any r on the right of the marble. For light, the illumination from the left represents an axial force towards the right. Far beyond 2f, the right hand background illumination fills in the umbra as a force to the left. The background forces from the left and right cancel where r is zero and 2f, and only the repulsion of the source of light in the marble is active. Near r equal zero, the force field is divergent and to the right, and appears to be repulsive from the left. Near r equal to but less than f, the force field is convergent and appears to be attractive from the right.

The force of the background is not apparent when r is much less than f, even if B is equal to A.

This is the optical model for forces.

The optical model can be turned into a model of gravity by an interpretation of the two coefficients, A and B.

Let A = -G* $M_\odot$ *$m_p$ and B = -G* $M_y$ *$m_p$.

G is the gravitational constant,

$M_\odot$ is the mass of the Sun,

$m_p$ is the mass of Pioneer 10,

$M_y$ is the effective mass of the background through the Sun.

f is found from the bending of light at the limb of the Sun. x is now the Solar gravitational force from Newton's law of gravitation, and y contains the model for the anomalous acceleration, $a_p$, of Pioneer 10. At some efficacious value of r, $M_y$ may be evaluated. With B = AR, $M_y$ is expressed as a fraction, R, of $M_\odot$ .

$$R(r) = \frac{m_p a_p}{y_r} \qquad (2.5)$$

and is independent of $m_p$, and is evaluated at some special r which is indicated by the data.

The model's one free parameter, $M_y$, is like $M_\odot$ in Newton's law. In any model, these types of constants are determined by data in experiments. In this sense an *ab initio* calculation of the anomalous acceleration of Pioneer 10 is performed. The model is not arbitrary, it is fixed by first principles.

The optical force for the gravitational focusing of gravitational force through the Sun can be expressed then by this formula:

$$F_0 = G \cdot M_\odot \left\{ \frac{-1}{r^2} - R \left[ \frac{1}{\left(f-r\right)^2} - \frac{1}{f^2} \right] \right\} \qquad (2.6)$$

Later, this can be integrated into the form of a potential.

Attraction and repulsion are interchanged in going from rays of light to rays of gravity. For r near the Sun the force is divergent and attractive to the left. Where r is less than and near f, the force is to the left and appears to be convergent and repulsive from the right. There is a local minimum in the radial force at f with spherical aberration.

In later sections the model is compared to the data, and the thesis of a constant additive acceleration, $a_p$, is examined, and compared to the optical model, and observational and experimental implications are delineated.

**Section 3:** Data Analysis

The optical model requires that two parameters be fixed for it to render predictions; they are: f and R.

f, the focal length of the Sun, may be found from the focal length of light in its bending around the limb of the Sun with the impact parameter, which is the radius of the Sun, the radius which is about 6.95 E+05 kilometers, need not actually be known, although it is. The angle subtended at the Earth at 1 AU by the Sun is about thirty minutes of arc, the half angle is fifteen minutes, some nine hundred seconds. The half angle subtended by the Sun at the focus, by virtual angles, is the same as the angle of bending, 1.75 seconds of arc.

The impact parameter, b, is the radius of the Sun, $R_0$, therefore:

$$R_\odot = b = 1 AU \cdot Tan(900 Sec) = f \cdot Tan(1.75 Sec) \qquad (3.1)$$

Using the small angle approximation and solving for f yields f = 514 AU.

R, the ratio of the effective external mass behind the Sun to the Sun's mass is selected in the following manner. A feature of the optical model that is important over other models is that the additional force is nearly zero in the inner Solar System, and does not appear to turn on until Pioneer 10 has reached the outer regions, where the anomalous acceleration has become evident.

Figure 7 in Anderson's analysis of Pioneer 10 data (Anderson, 2002) shows the turn on of the anomalous acceleration as it emerges and becomes evident over radiation effects from the Sun. The shape of the curve's turn on in this figure is reminiscent of a D vs. Log(E) curve in photography, or the S-Sigma curve in project management. It has a base leading into a toe, then rises and reaches a saturation phase, and thereafter is approximately constant, or some slight curve.

Figure 8, also from Anderson's analysis of the Pioneer data (Anderson, 2002) shows the speed of Pioneer 10 as a downwards sloping curve with annual undulations and glitches, corresponding to an on the average, constant negative radial acceleration. A broad sloping curve that lies within the data is a possibility.

Until Pioneer 10, there was no directly measurable need of a nuance to gravitation within the Solar System, and while there was no problem for which the optical model could be used, it now is the simplest reasonable model.

R(r) is chosen to be at the value of r at the toe where the rise begins, which is at 10 AU, and the data analysis continues from the saturation value of 13 AU. R = 0.97 at 10 AU, which is to say, that $M_y = 0.97$ $M_\odot$, and R is about 1. f and R are now fixed without a least squared parameterized fit, entirely by the physics of the problem.

Data for Pioneer 10 for r was downloaded from the NASA website for Pioneer 10 (NASA 2004). The data from NASA (2004) averaged over days is r in AU versus t as date. The first difference is proportional to the speed, and the second difference is proportional to the acceleration, a(r). Although the averages, and the differences are offset, no attempt has been made to modify or correct them.

Most differences are zero, those which are not, were used. A four sigma elimination for fliers has been applied. The data for r ranges from 13 - 74 AU. The set of second differences represent accelerations versus positions. Ratios of accelerations to a force model are independent of an overall proportionality constant, so long as it is not so small as to interfere with the precision of numerical calculations. It appears that the reduced data has a great deal of scatter in it, but something is better than nothing.

The method used to compare the Data and results to Theory is the goodness of a linear regression fit to their ratio, point by point, Data/Theory ( D/T ).

There are two competing theories for the anomalous acceleration of Pioneer 10:

1) x - $a_p$, and 2) x + y.

Recall that x is Newtonian gravitation, $a_p$ is an assumed constant acceleration, and that y is the additional term from the optical model which is based on first principles.

If one or the other of the theories is manifestly incorrect, then the coefficient for D/T as a straight line will be smaller, if the theory is correct, then the coefficient for D/T as a straight line will be larger; or in other terms, they each have a linear regression coefficient, and one may be better than the other.

The first trial regression is for a straight line to the reduced data of the second difference. This fit tests the data versus no theory, and if any theory is worse than this, it or the data are wrong. The regression coefficient for a straight line fit to the data is; $r^2 = 0.03$. The fit is almost perfectly poor, which would be 0.00. Certainly some theory will do better than this.

For Theory-1, Data - a(r) / (x(r) - $a_p$) on the same data set yields a value of another value for $r^2$, which is larger and therefore better than nothing.

For theory-2, Data - $a_r$ / ( x(r) + y(r) ), again on the same dataset, yields the same linear regression coefficient as the other. This too is better than nothing, and it shows that Theory-1 and Theory-2 have the same significance.

Theory-1 / Theory-2, their ratio point by point, varies at most by 0.2% in the range of 13 - 74 AU. Their significances by the closeness of their ratios, point by point, should be the same, and they are. The models are equivalent except for their observational implications.

**Section 4** Observational and Experimental Implications

4.1) Integrated Background

Having come to the realization that the effective background mass, $M_y$, that passes gravity through  the Sun is almost exactly the same as the Solar Mass, $M_\odot$, it is interesting to find its value over a sphere centered on the Sun, without credence to the fact that it is only a part of the galactic and cosmic image formed by the marble of the Sun. Using that $M_y$ subtends and angle across the Sun, and that it is twice the bending angle of 1.75 seconds, and so it is 3.5".

Given that the circumference, (C), of a great circle around a sphere and that the area, (A), of the sphere are related to its radius, (r), the area can be written in terms of the circumference.

Using:

$C = 2\Pi r$ and $A = 4\Pi r^2$

then

$A = C^2/\Pi$

In the sense that there are $2\Pi$ radians in a circle laid out on its circumference, the formula yields that there are $4\Pi$ steradians laid out on the surface of the sphere.

That is:

$A = (2\Pi \text{ Radians})^2/\Pi = 4\Pi$ Steradians

In the same sense, the number of st-seconds are computed on the surface of a sphere, and also the number of $M_y$, which subtend 3.5 seconds onto the surface of the sphere, and this constitutes the method of integration.

Given that there are 1.296 E + 06 seconds subtended by the Circumference, there are then about; 5.34 E + 11 st-seconds on the surface of the sphere. Appling this technique to $M_y$, equal to $M_\odot$, which subtends 3.5", the integrated background is 4.26 E + 10 $M_\odot$. This means that the integrated background is about 40 billion Suns. This value is about 1/10th the mass of the local galaxy of the Sun.

4.2) Evaluation of the Background

This value arises from the matter on a line parallel to the motion of Pioneer 10 and 11, and the tangential trajectory of the Sun in its motion about the core, that is perpendicular to the direction of the galaxy's core, and it is exclusive of the greater mass in the core. The small angle of the focus, is in part a function of the large focal length, and is in part responsible for the large integrated background. The real effect driving to such a large background is the amount of matter in line with the trajectory of Pioneer 10.

The matter on a tangent to the Sun's motion, cuts across the outer third of the galaxy, and the four PI integration adds a fictitious amount of mass above the ecliptic, but there is not that much matter beyond the Sun's orbit. Shades of Olber's paradox are indicated, the cosmic background is in part responsible for the integrated background. This is consistent with the spacing of the galaxies between their cores and rims. There apparently is a cosmic background of considerable size.

4.3) Effects at f and 2f

It happens that the location of the Kuiper Belt is near the value for f. It is also happenstance that the location of the Oort Cloud is at 2f to nf. These are cases of clouds bound in local minima of focused gravitational forces from behind the Sun.

It happens that at 2f, the rays of external matter focused thought the central object are cancelled in their effect, and matter on opposite sides of a central mass are attracted to each other by focusing through the center. The center with a focal length of f, focuses rays from 2f on the left to a region at 2f on the right, and left and right, each at 2f from the center are mutually attractive.

There are two obvious examples of interest in known galactic structures. They are Hoag's object (See Figure 1) and the Starburst galaxy NGC 7742 (See Figure 2.)

For a circular system such as Hoag's object, it can be argued, that the core is of radius f, and the ring or rim around it is of radius 2f. The core which extends to a radius of f, provides the focusing for the ring, everywhere at 2f, to be bound by its own mass about the central core.

The same argument applies to the Starburst galaxy.

These two galaxies both appear to satisfy this arrangement of core and ring structures at f and 2f.

Objects within 2f of each other with a bright gravitational background, are inside each other's umbra. This situation gives rise to an apparent repulsive force between them, although this repulsion is the attraction of the background pulling them apart. This mechanism must apply to clusters as well, and further, matter between them, such as, the core and the rim, is swept away and towards each of them.

This feature of sparse populations between f and 2f is also exhibited by the sample galaxies.

4.4) Neutron Stars

Imagine a neutron star, which has as its mass, one solar mass; i.e., $M_n = M_\odot$. A reasonable value for its radius is about ten kilometers, which is 1/69,500th of $R_\odot$, that of the Sun. By proportions then, its focal length, given the same mass as the Sun, but with a smaller radius and impact parameter, is smaller by the same amount. Using that the radius of the Sun, $R_\odot = 6.95 \text{ E} + 05$ kilometers, and that the focal length of the Sun, $f_\odot$, is 514 AU, and also that, an AU is $1.495 \text{ E} + 08$ KM, then the focal length of this kind of neutron star is; $f_n = 1.5 \; R_\odot$.

Given that there is a focal length of a Solar mass neutron star, it is reasonable to suppose that, like the Sun with a Kuiper belt and Oort cloud set of structures around it, that also the neutron star will exhibit proportionally the same types and kind of structures around it at its $f_n$ and also $2f_n$.

These structures about a neutron star are the result of transmission of the gravitational fields of stellar partners and its galactic and cosmic background through the neutron star with matter of approximately constant nuclear density, with positive spherical aberration.

To some degree as observational technique progresses, these features will be delineated and catalogued.

4.4) Integration of the forces to potentials

The model with a constant acceleration can be integrated to a potential, $V_j$ per unit mass:

$$V_j = G \cdot M_\odot \frac{-1}{r} + a_p \cdot r \qquad (3.2)$$

The optical model can be easily integrated to the form of a potential, $V_o$ per unit mass:

$$V_o = G \cdot M_\odot \left\{ \frac{-1}{r} + \frac{R}{(f-r)} - \frac{R \cdot r}{f^2} \right\} \qquad (3.3)$$

Both $V_k$, ( k=j,o ) can be inserted from the weak field approximation into the metric, with signature, (-+++).

$$g_{00} = -\left( 1 + \frac{2V}{c^2} \right) \qquad (3.4)$$

Using the conserved quantity for light;
$$g_{00}(r) \cdot E^2(r) = g_{00} \ (r\text{'}) \cdot E^2(r\text{'}) \qquad (3.5)$$

where $E = h*f$. It is possible to imagine that the $+ a_p *r$ term in $V_j$ encourages an infinite red shift to black as r' increase without limit, yet the light of distant stars is visible, and so the, $a_p*r$ term is unphysical at great distances. In addition, rays that are bent from around the Sun will exacerbate the effect.

While, $V_0$, contains a term that is, $-r/f^2$, from the umbra, it is certain that rays from around the Sun will wash it away for r much larger than f.

The first term of the additional optical model for the potential per unit mass; namely,

$$\frac{1}{(f-r)}$$

is left after cancellation from the isotropic background. For r near and larger than f, $g_{00}$ develops a singularity, from:

$$g_{00} = -\left(1 + 2 \cdot G \cdot M_\odot \cdot \frac{R}{c^2 \cdot (f-r)}\right) \qquad (3.6)$$

Solving for r:

$$r_{sf} = 2 \cdot G \cdot M_\odot \cdot \frac{R}{c^2} + f \qquad (3.7)$$

The first term on the rhs is the Schwarzschild radius of the background that is focused through the Sun, and the second term is the focal length of the Sun.

In the same sense that it was wondered if the Schwarzschild radius was real, now wonder if $r_{sf}$ is real. It appears that for r slightly larger than f, there may be an event horizon. This is a region where if light is emitted, it will be red shifted, and in this zone light may orbit the Sun. Like Alexander's dark band in rainbows, there may be a dark zone for objects emitting light outwards from stellar systems.

This all assumes that gravity is bent by the Sun and that spherical aberration is minimal.

Now the relative shifting of the color of in coming light on the optical axis of an observer and the Sun can be calculated.

$$z = \left| g_{00}(r) \right|^{-\frac{1}{2}} - 1 \qquad (3.8)$$

This is expected to yield a red shift of incoming light within the umbra, except on the optical axis, where the color shift is expected to be towards the blue.

Given that an observer on the Earth with a spectroscopic telescope can track the 15 arc seconds of angle swept by the motion of the Earth per second of time, there are ample opportunities to observe stellar objects which have incurred a blue shift within the field of view. Centered on the umbra, a red shift may be viewed, followed by a greater blue shift closer to the precise optical axis.

F) Pull on Test Masses

As a test particle, the space probe, approaches the focal point, it is evident that what was expected to be an adequate escape speed from the Solar System, is perhaps too small. The situation is reminiscent of the breakers around an island which can prevent a small row boat from escaping the island.

The pull on a test particle coasting away from the sun will act like the breakers around an island on a bottle impeding its outward motion, and repelling it back inwards towards the Sun.

**Section 5:** Conclusions

The anomalous acceleration of Pioneer 10 has been examined via *ab initio in vacuo* calculations according to first principles. There are no free parameters, and so the results are a matter of mechanics. It must be so. The resulting effects, while they remain discussable, are essentially *a priori* analytic.

There must be the bending of gravitational forces through the Sun, they have focal properties that effect the motion of Pioneer 10, despite any other proposal that may be suggested to account for the same effects.

There are magnified effects of gravity for a test mass approaching the focus. There is an effective background mass, $M_y$, which is about the same as that $M_\odot$. They add together to apparently increase the force regime of the Sun in distant regions.

Structures such as the Kuiper belt and the Oort cloud naturally appear with a separation between them; and this structural affect ramifies to neutron stars and also to galaxies.

The inescapable observation is that within stellar systems such as the Solar system and galaxies, is that, they have the same structures with tenuous gaps. Stars have clouds with a gap, and galaxies have cores with rims or rings. The affects at f and 2f seem to obviate a need for some dark matter.

In addition to these observational effects, there are implied color shifts to incoming light from outside sources that can be experimentally verified.

The optical model is consistent with known physics and by implication reproduces known observational effects.

This work represents a first approximation to an ideal model.

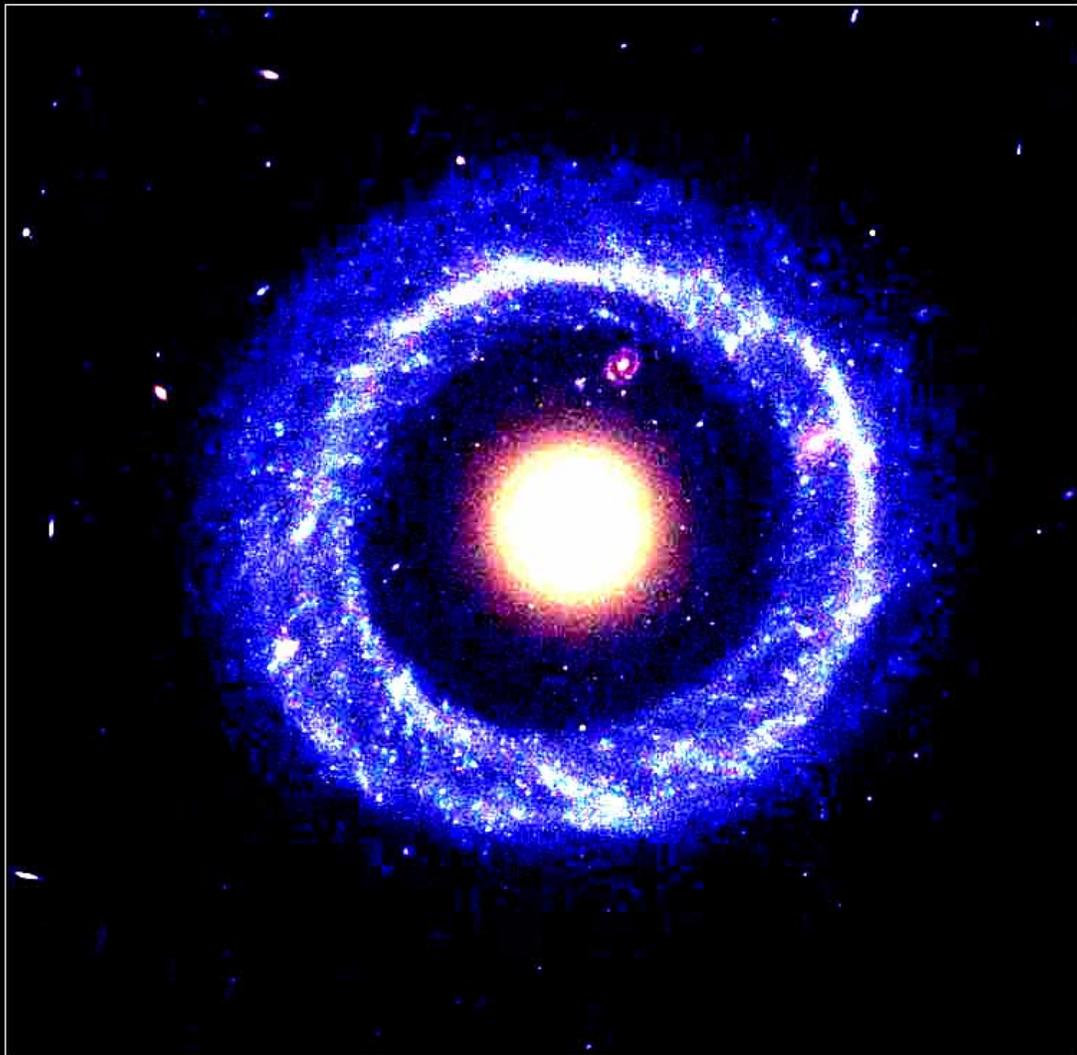

Figure 1 (Hubble Heritage, 2005)

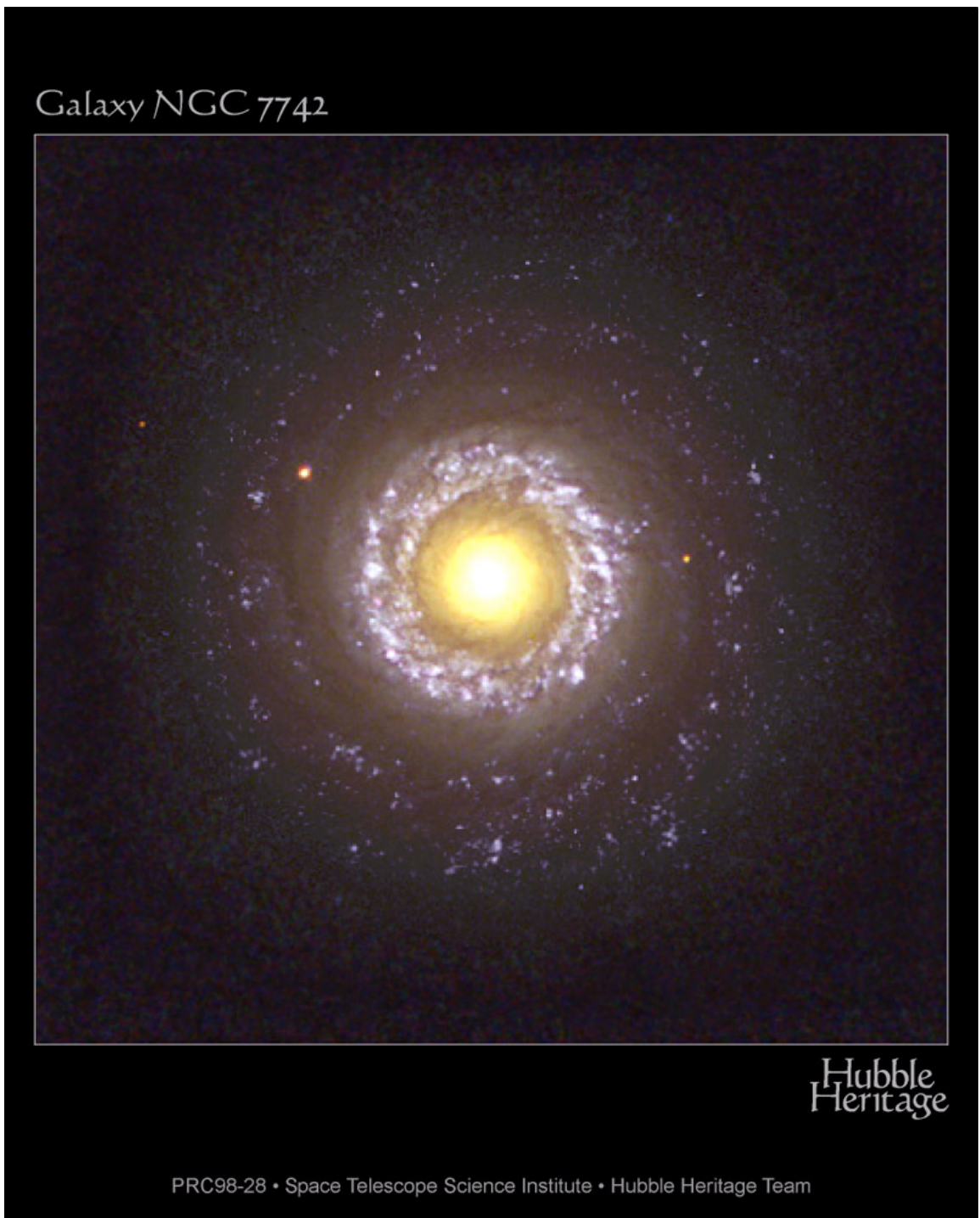

Figure 2  (Hubble Heritage, 2005)

List of Figures: